\documentclass[11pt,twoside]{article} 
\usepackage{asp2004}
\usepackage{epsf}
\usepackage{psfig}
\usepackage{lscape} 

\begin{document} 
\title{How Many Hot Subdwarf Stars Were Rejected from the PG Survey?}
\author{R. A. Wade,$^1$ M. A. Stark,$^1$ and R. F. Green$^2$} 
\affil{$^1$Pennsylvania State University, Dept. of Astronomy \& 
Astrophysics, 525 Davey Lab, University Park, PA 16802, USA\\
$^2$ NOAO, P.O. Box 26732, Tucson AZ, 85726-6732, USA}
%
\begin{abstract}
   It has been suggested that many hot subdwarfs lurk in the pile of
rejected UV-excess candidate stars from the Palomar-Green (PG) survey.
This suggestion is not supported by available photometric data.
\end{abstract}

\section{Introduction}

During the PG survey for ultraviolet excess (UVX) objects, candidate
UVX objects were those with (transformed) $U-B < -0.46$.  It was
recognized that the large error in $U-B$, $\sigma\approx 0.38$, meant
that color selection should be supplemented by spectroscopy for
classification, since more accurate temperature information was likely
available from the spectra than from $U-B$. Many {\em candidate} UVX
targets were indeed culled from the final PG catalog (Green et al.
 1986: GSL86), because their classification spectra showed
the Ca II K line in absorption. These K-line stars were thought not to
be genuinely hot, but rather to be metal-poor subdwarf F or G stars
that crept into the candidate list owing to a combination of low
metal-line blanketing and photometric errors.

Hot subdwarf stars (especially sdB stars, but including some sdO
stars) are understood to belong to the Extended Horizontal Branch
(EHB), and as such are core-helium burning objects with very thin
hydrogen envelopes. Recently, there has been renewed interest in
scenarios of the origin of hot subdwarf stars that involve binary star
processes (Roche-lobe overflow, common-envelope evolution) to strip
the hydrogen-rich envelope away from the helium core, near the time of
He-core ignition (Han et al.\ 2002, 2003).

In comparing their population synthesis models with observations, Han
et al.\ (2003) drew attention to the K-line stars rejected from the PG
catalog, as possibly representing a ``missing'' group of hot
subdwarfs, hidden by their cooler and (somewhat) brighter binary
companions.  Owing to dilution of the hot star's energy distribution
by the companion, the $U-B$ color could be marginal for the PG color
criterion, while the cool star would contribute a K line, much as a
metal-poor subdwarf would show.  In this interpretation, therefore,
these ``PG--rejects'' actually belong in the PG catalog, and moreover
would constitute important evidence in favor of binary formation
channels for sdB.

Han et al.\ have put forward a hypothesis that can be tested.  The
list of PG--rejects exists in a card file (with finding charts) kept by
RFG.  Here we report our investigation to date of the PG--rejects.

\section{The Sample of Rejected K-line Stars}

We have assembled a catalog of 1125 distinct PG--rejects.  We found
291 stars that are present in both the Two-Micron All-Sky Survey
(2MASS) Point Source Catalog and the Sloan Digital Sky Survey (SDSS)
DR2 survey region.  Here we focus on the 173 stars with Sloan $r$
magnitudes in the range 14.00 to 16.00 (median $r=14.86$). Of these,
136 have all 5 optical magnitudes ($ugriz$), and all were measured in
the 2MASS $J$, $H$, and $K_s$ bands.

Many K-line stars were observed spectroscopically {\em before} the
final $U-B$ transformation (from photographic to Johnson $U-B$) was
established.  In the subsample of 291 stars, only 150 would
have met the final $U-B$ criterion for inclusion in the PG catalog,
while 131 have final $U-B$ colors redder than the catalog cutoff and
10 were from survey fields not included in the final catalog.
\begin{figure}[!ht]
\resizebox{\textwidth}{!}{\plottwo{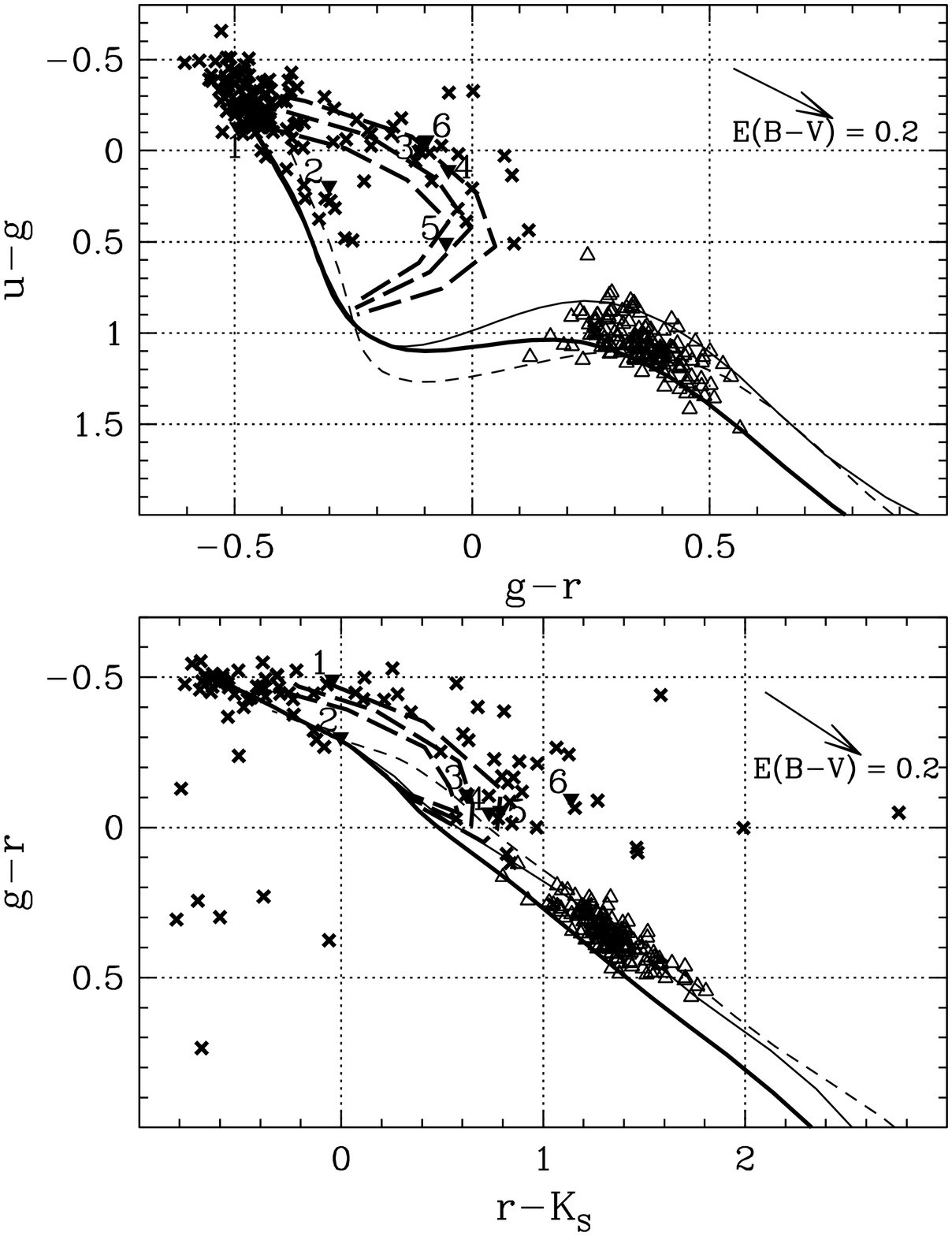}{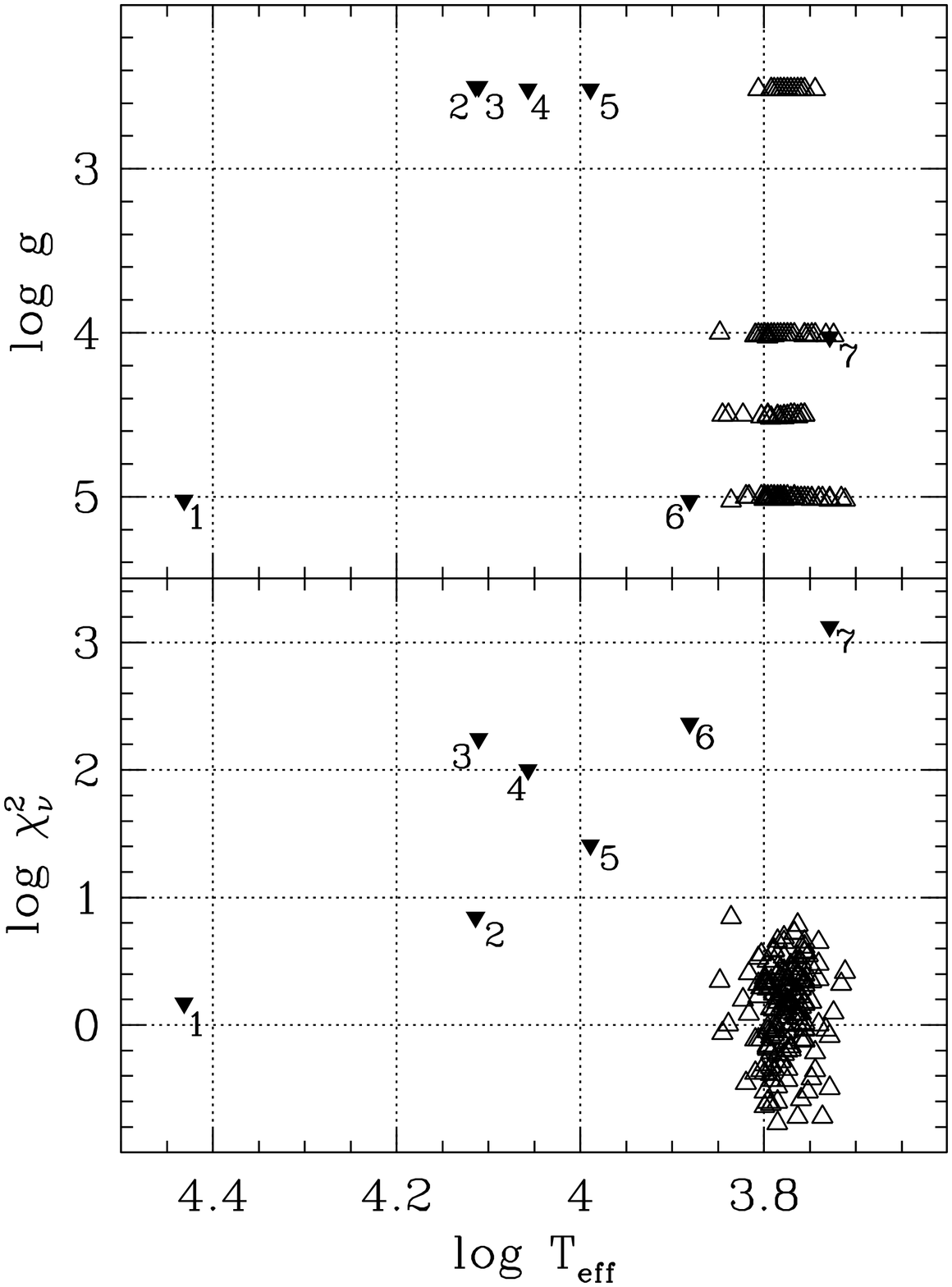} }
\caption{{\em Left panels:} Color-color diagrams of hot subdwarfs from the PG catalog
(crosses) and K-line ``PG--reject'' stars (triangles).  Errors on $K_s$ are
sometimes large for the PG subdwarfs, contributing to the apparent
scatter in the lower panel.  Heavy black line -- main sequence locus,
$\log g = 4.5$, solar metallicity [M/H]=0.0; thin black line --
metal-poor main-sequence locus, $\log g = 4.5$, [M/H]=$-1.5$; thin dashes line,
metal-poor giant locus, $\log g = 2.5$, [M/H]=$-1.5$.  heavy dashes --
composite colors for hot subdwarf + main sequence binaries.
{\em Right panels:}  Results of fitting single-star models to the SDSS/2MASS
photometry of PG--reject stars.  Seven objects are clear outliers
either in temperature or in the quality of the fit (filled triangles),
while 166 stars have satisfactory fits and are tightly clustered (open
triangles). 
{\em Upper:} $T_{\rm eff}$ and $\log g$. Objects shown at low $\log g$
may have fits at higher $\log g$ that are only slightly worse.
Stars with lower metallicity [M/H] are plotted with a slight offset
downward; otherwise $\log g$ has only four discrete values.
{\em Lower:} $\chi^2_\nu$ is
satisfactory (smaller than $\sim$10) for 166 stars.  Median
value is $\chi^2_\nu=1.49$.  Further improvement is possible
by using additional values of [M/H] and $\log g$, plus
interstellar reddening.}
\end{figure}

Figure 1 shows aspects of the SDSS/2MASS photometry for the 173 K-line
stars (rejected from the PG catalog), along with data for 199 PG stars
classified as hot subdwarfs.  Not all of the objects for which
SDSS/2MASS photometry is available are shown in each diagram, i.e.,
sometimes the needed color cannot be computed.  Also shown in the
Figure are loci for the main sequence (solar abundance), metal-poor
main sequence, and metal poor giants representing horizontal branch
stars.

Three sequences of composite (binary) models are also shown.  These
represent the light from a hot subdwarf star ($T_{\rm eff} = 25000$~K,
30000~K, and 35000~K, with $M_V$ derived from the zero-age EHB
calculations of Caloi 1972), combined with the light from a cool
main-sequence companion (eight $T_{\rm eff}$'s ranging from
4000~K to 9750~K).  These sequences emerge from the hot end of the
stellar locus (faintest, coolest companions at this end), loop
away from the single-star locus, and then loop back to meet the
stellar locus at a (single-star) $T_{\rm eff}$ near 10000~K.

The recognized PG hot subdwarfs and the K-line PG--rejects are very
different groups of stars.  In this sense, the classification
spectroscopy carried out by GSL86 was successful in improving on the
photographic $U-B$ color selection.  It is also clear that some PG
subdwarfs are composite objects (a conclusion already reached by Stark
\& Wade 2003; see also Reed \& Stiening 2004, and poster by Stark \&
Wade at this conference).  Finally and most important, {\em the
PG--reject stars are consistent with being single stars}, possibly
metal poor, just as they were interpreted to be by GSL86.  Except for
a few outliers, {\em they are not binaries composed of a ZAEHB
subdwarf and a main-sequence companion.}

\section{Fitting the PG--reject Stars as Single Stars}

We compared the observed magnitudes for the 173 stars with model
magnitudes, derived from the synthetic photometry done by the Padova
group (Girardi et al.\ 2002, 2004) using Kurucz stellar atmosphere
energy distributions. 
As single-star models for the PG--reject stars, we considered all
available Padova models with $T_{\rm eff}$ in the range 4000 -- 50000~K
($\log g = 4.0$, 4.5, 5.0), with metallicities [M/H] = 0.0,
$-1.0, -1.5, -2.0, -2.5$.  We interpolated in $T_{\rm eff}$ to make the grid
denser.

For each of the 173 stars, we scaled each model in brightness to
find the best fit.  We chose as the best overall model, that model
that gave the smallest reduced chi-square statistic, $\chi^2_\nu =
\chi^2/{\rm dof}$, where ${\rm dof} = 6$ (or 7) is the number of
degrees of freedom for 7 (or 8) valid magnitudes.

The main results are summarized in Figure 1 (right--hand panels).
Seven outliers have either large $\chi^2_\nu$ or unusual $T_{\rm
eff}$ or $\log g$. All of the remaining 166 PG--rejects are fitted
with $T_{\rm eff}$ in the range 5000 -- 7100~K. Most of the stars (136
of 166) are preferably fitted with low-metallicity models, [M/H] =
$-1.0$ or below. These are {\em consistent with the GSL86
interpretation that these are metal-poor F and G subdwarfs}.  The
locus for low-gravity models in Figure 1 also passes through the
cluster of PG--reject stars, so sometimes the absolute best-fitting
model by the $\chi^2$ criterion is a low-gravity model (39 cases out
of 166); models nearly as good will likely be found at higher $\log
g$.

Given our present understanding of the systematics of the SDSS error
estimates and the incompleteness of our model grid, the $\chi^2_\nu$
values for the 166 non-outliers are acceptably small.  A trend in
$T_{\rm eff}$ with $r$ magnitude indicates that reddening may need to
be taken into account for the fainter (more distant) stars; when this
is done, the $\chi^2_\nu$ values may decrease further.

Some outliers are identified by numbers in Figure 1.  Two of the
outliers have SDSS spectra.  {Star \#2} may be a blue horizontal
branch star.  {Star \#3} appears not only in the PG--reject list, but
also in the PG catalog itself (PG 1723+603)! The spectrum shows Mg Ib,
Na D, and Ca II infrared triplet absorption, but the continuum is
blue. In Figure 1, it lies in the region of the composite models.
{Star \#5} also lies in the region of the composite models in Figure 1
with similar but weaker evidence of `cool' stellar features such as Mg
Ib and Na D.  The spectral energy distribution of {Star \#7} shows
`excesses' at both the short and long-wavelength ends relative to the
(not-so-good) best-fitting single-star model, suggesting that it is
composite.  {The other outliers} lie either close to the hot
single-star locus or close to the sequences of composite models.

\section{Summary}

A very few objects in our sample of PG--reject stars may plausibly be
binary systems with a hot subdwarf star component.  Also, a few
objects seem to have entered the PG--reject list by accident. The vast
majority of the PG--reject stars, however, are sufficiently modeled as
single stars consistent with their being the metal-poor sdF/sdG
contaminants that GSL86 were guarding against. The color-color
sequences of sdB + cool star binaries (with $M_V$ for main sequence
companions!)  are well separated from the observed colors of the
PG--reject stars.  There is at present no compelling evidence for
large numbers of additional hot subdwarf stars hiding in binaries that
were rejected from the PG catalog.

\acknowledgements{We acknowledge helpful discussions with P.\
Durrell, C.\ Gronwall, and R.\ Ciar\-dullo.  We made use of SDSS Data
Release 2, the 2MASS, the USNO-A2 catalog, and SIMBAD.  Supported by
grants from NASA.}

\end{document}